\documentclass[journal,twoside]{IEEEtran}
\usepackage[hyphens]{url}
\usepackage{hyperref}
\hypersetup{breaklinks=true}
\usepackage[square, comma, sort&compress, numbers]{natbib}
\usepackage{amsmath}
\usepackage{amsfonts}
\usepackage{amssymb}
\usepackage{algorithm}
\usepackage{algorithmic}
\usepackage{paralist}
\usepackage{graphicx}
\usepackage{subfigure}
\usepackage{booktabs}
\usepackage{multirow}
\usepackage{makecell}
\usepackage{multicol}
\usepackage{paralist}
\usepackage{enumerate}
\usepackage{amsthm}
\usepackage{bm}
\usepackage{array}
\usepackage{psfrag}
\usepackage{eurosym}
\usepackage{blkarray}
\usepackage{threeparttable}
\usepackage{extarrows}
\usepackage{color}
\usepackage{bpchem}           
\usepackage{textcomp}  
\usepackage{placeins}   
\usepackage{booktabs}

\usepackage{colortbl}	

\usepackage[OT1]{fontenc}  

\newtheorem{Lem}{Lemma}

\newtheorem{Def}{Definition}

\newtheorem{Rem}{Remark}

\graphicspath{{./eps/}}

\definecolor{mygray}{gray}{.9}

\begin{document}
%
\title{Efficient Channel Estimator with Angle-Division Multiple Access}%
%
%
\author{Xiaozhen~Liu,
        Jin~Sha,
        Hongxiang~Xie,
        Feifei~Gao,
        Shi~Jin,
        Zaichen~Zhang,~\IEEEmembership{Senior~Member,~IEEE},
        Xiaohu~You,~\IEEEmembership{Fellow,~IEEE}
        and~Chuan~Zhang,~\IEEEmembership{Member,~IEEE}\\
\thanks{Xiaozhen Liu and Jin Sha are with the School of Electronic Science and Engineering, Nanjing University, China. Hongxiang Xie and Feifei Gao are with the Department of Automation, Tsinghua University, Beijing, China. Shi Jin, Zaichen Zhang, Xiaohu You, and Chuan Zhang are with the National Mobile Communications Research Laboratory, Southeast University, Nanjing, China. Email: shajin@nju.edu.cn, chzhang@seu.edu.cn.}
\thanks{This paper was presented in part at International Conference on ASIC (ASICON), Guiyang, China, 2017 \cite{liu2017vlsi}. \emph{(Corresponding author: Jin Sha and Chuan Zhang.)}}
}

%
%

\markboth{}%
{X. Liu \MakeLowercase{\textit{et al.}}: Efficient Channel Estimator with Angle-Division Multiple Access}
%



\maketitle

\begin{abstract}
Massive multiple-input multiple-output (M-MIMO) is an enabling technology of 5G wireless communication. The performance of an M-MIMO system is highly dependent on the speed and accuracy of obtaining the channel state information (CSI). The computational complexity of channel estimation for an M-MIMO system can be reduced by making use of the sparsity of the M-MIMO channel. In this paper, we propose the hardware-efficient channel estimator based on angle-division multiple access (ADMA) for the first time. Preamble, uplink (UL) and downlink (DL) training are also implemented. For further hardware-efficiency consideration, optimization regarding quantization and approximation strategies have been discussed. Implementation techniques such as pipelining and systolic processing are also employed for hardware regularity. Numerical results and FPGA implementation have demonstrated the advantages of the proposed channel estimator.
\end{abstract}


\section{Introduction}\label{sec:intro}
\IEEEPARstart{W}{ith} the explosive growth of mobile applications, cloud synchronization services and the rapid development of high-quality multimedia services such as high resolution image and 4K-resolution high dynamic range (HDR) video streaming, the existing 4G mobile communication technology could not meet the needs of enterprises and consumers for wireless communication networks any more. As a result, 5G mobile communication technology \cite{andrews2014will,shafi20175g,rusek2013scaling} has been raised up with higher transmission speed, stronger bearing capacity, and a wider range of applications. Massive multiple-input multiple-output (M-MIMO) is one of the key technologies of 5G \cite{boccardi2014five,larsson2014massive}, owing to its plenty advantages, such as high spectral efficiency, high power efficiency, and high robustness. The performance of M-MIMO systems relies heavily on the acquisition of the uplink (UL) and downlink (DL) channel state information (CSI). However, the large-scale antenna array of M-MIMO brings new challenges to channel estimation \cite{xie2016overview} :

\begin{itemize}
	\item The overhead of training sequence grows due to the increasing number of users while the reuse of training sequence will arouse pilot contamination \cite{marzetta2010noncooperative,jose2009pilot,fernandes2013inter,you2015pilot}.
	\item The growing dimension of channel matrices (CMs) or channel covariance matrices (CCMs) makes the complexity and resource consumption of the traditional UL and DL channel estimation algorithm greatly increased, limiting the M-MIMO system to play its superiority.
	\item The amount of CSI that users feed back to the base station (BS) during DL channel estimation is growing with the increase of the number of antennas at the BS, which is a great burden of the feedback channel.
	\item Channel reciprocity makes it easy to acquire DL CSI from UL CSI for time division duplexing (TDD) systems, while the non-reciprocity characteristic causes that the DL channel estimation of frequency division duplexing (FDD) systems cannot be predigested, which is a great burden for user's devices.
\end{itemize}

In order to reduce the computational complexity, we need to take advantage of the low-rank properties of the channel, which can reduce the dimension of CMs and CCMs significantly and acquire the valid information. Many works point out that the directions of arrivals (DOA) as well as the directions of departures (DOD) of propagation signals are limited in a narrow region (i.e., the angle spread (AS) is small) because the BS equipped with large-scale antenna array has to be located on the top of high buildings, which is known as the finite scattering model \cite{burr2003capacity}. Another similar scenario is the mmWave communication, where the channel is naturally sparse and the AS equals $0$ \cite{gao2016energy}. Meanwhile, due to the large-scale antenna array of M-MIMO, the spatial resolution of the BS is significantly improved, which means the BS can distinguish users from different directions more easily so that the representation of channel can be strongly sparse and there are relatively few non-zero elements in CMs and CCMs. As a result, a lot of new or optimized methods to acquire CSI have been proposed \cite{adhikary2013joint,nguyen2013compressive,rao2014distributed,sun2015beam,fang2017low}, especially for DL channel estimation of FDD system due to its non-reciprocity characteristic. \cite{adhikary2013joint} proposed an approach under joint spatial division and multiplexing (JSDM) scheme for DL channel estimation of FDD system, where the sparsity of CCMs is exploited and the eigenvalue decomposition (EVD) algorithm is required, which is a challenge for implementation.
\cite{nguyen2013compressive} proposed a low-rank matrix approximation
based on compressed sensing (CS) and solved via a quadratic semi-define programming (SDP), which is novel but far too complex to implement. \cite{rao2014distributed} deployed a CS-based method with the joint channel sparsity model (JCSM) which utilizes virtual angular domain representation of the CM and limited local scattering in order to reduce the training and feedback overhead. To this end, we first proposed a transmission strategy based on spatial based expansion model (SBEM) in \cite{xie2017unified}, which comes from array signal processing theory. This scheme is also known as angle-division multiple access (ADMA) scheme. ADMA scheme has some particular advantages:

\begin{itemize}
	\item Due to the increased angle resolution of antenna arrays at the BS, the angular information can be easily obtained by the discrete Fourier transform (DFT) of CMs under ADMA scheme.
	\item The angular information is corresponding to the real directions of users with the array signal processing theory, while the others' methods only have a virtual angular representation.
	\item As a result of the reciprocity brought by the DOA and DOD, the complexity and overhead of DL channel estimation and feedback can be reduced.
	\item The estimation algorithm mainly contains DFT calculation, matrix multiplication, sorting and grouping, which is convenient for implementation.
\end{itemize}

As it has shown in \cite{xie2017unified}, the performance of ADMA is better than \cite{adhikary2013joint} and \cite{rao2014distributed}, especially at low signal noise ratio (SNR). In addition, there are also blind and semi-blind estimation methods to be explored. Those methods have higher transmission efficiency because they need fewer (or no) training sequences. But the result of those methods may be not accurate at the start of transmission because the BS needs some time to accumulate channel statistics information. Moreover, the efficient implementation of ADMA is very challenging due to the non-linear computation involved in algorithm level, therefore hinder its application for channel estimation.

In order to bridge the aforementioned gap between algorithm and implementation, this paper devotes itself in proposing the hardware architecture for channel estimation under ADMA scheme for the first time. Hardware-aware partition of the algorithm is conducted. Our main technical contributions can be listed as follows:

\begin{itemize}
  \item We propose a hardware-efficient channel estimator under ADMA, which takes the advantage of the sparsity of M-MIMO systems in order to reduce the complexity, save the amount of training sequences, and speed up the channel estimation of large amount of users.
  \item We discuss the approximation of algorithm and transmission strategy as well as the quantization optimization in order to make the our channel estimator suitable for hardware implementation.
  \item We propose the first channel estimator architecture with ADMA scheme, successfully achieve higher hardware efficiency and higher processing speed for channel estimation of M-MIMO systems.
  \item We develop an optimized architecture to simplify our original channel estimator, with little performance loss but huge resources reduction and higher hardware efficiency.
  \item We present the FPGA implementation of our ADMA channel estimator on Xilinx Virtex-$7$ xcvu$440$-flga$2892$-$2$-e, to demonstrate its suitability for $5$G wireless. The advantages have been verified by FPGA implementations.
\end{itemize}


The remainder of this paper is organized as follows. Section \ref{sec:pre} proposes the implementation-aware partition of ADMA algorithm. The hardware-friendly approximation and the simulation results are presented in Section \ref{sec:sim}. The detailed pipelined hardware architecture is presented in Section \ref{sec:hard}. FPGA implementation is also given in the same section to demonstrate the advantages. Finally, Section \ref{sec:con} concludes the entire paper.

\textbf{\textit{Notations}}. \textcolor{black}{The notations} employed in this paper are listed in Table \ref{table:notation} for clearer representation.

\begin{table}[ht]
\centering
\caption{Notations in This Paper}
\begin{tabular}{c|l}
\Xhline{1.0pt}
Symbol&  \multicolumn{1}{c}{Definition}\\
\hline
\rowcolor{mygray}
$M$ & number of antennas at the BS, \\
$K$ & number of users that the BS serves,\\
\rowcolor{mygray}
$L$ & length of training sequences,\\
$\tau$ &number of parameters the BS can handle,\\
\rowcolor{mygray}
\hline
$\mathbf{h}$ / $\mathbf{H}$ & vector $\mathbf{h}$ / matrix $\mathbf{H}$,\\
$\left[\mathbf{h}\right]_i$ & the $i$-th element of vector $\mathbf{h}$,\\
\rowcolor{mygray}
$\left[\mathbf{H}\right]_{ij}$ & the $(i,j)$-th element of matrix $\mathbf{H}$,\\
${\mathbf{h}}^T$ / ${\mathbf{H}}^T$ & the transpose of vector $\mathbf{h}$ / matrix $\mathbf{H}$,\\
\rowcolor{mygray}
${\mathbf{h}}^H$ / ${\mathbf{H}}^H$ & the Hermitian of vector $\mathbf{h}$ / matrix $\mathbf{H}$,\\
$\cal B$ & set $\cal B$ of $\tau$ continuous integers,\\
\rowcolor{mygray}
$\left|\cal B\right|$ & the cardinality of the set $\cal B$,\\
$[{\mathbf{h}}]_{\cal B}$ & sub-vector of $\mathbf{h}$ by keeping the elements indexed by $\cal B$,\\
\rowcolor{mygray}
$[{\mathbf{H}}]_{:,{\cal B}}$ & sub-matrix of $\mathbf{H}$ by collecting the columns indexed by $\cal B$,\\
${\rm{diag}\left\{{\mathbf{h}}\right\}}$ & \makecell[lc]{a diagonal matrix with the diagonal elements constructed \\ from vector $\mathbf{h}$,}\\
\rowcolor{mygray}
${\mathbb{E}}\left\{\cdot\right\}$ & the statistical expectation.\\
\Xhline{1.0pt}
\end{tabular}
\label{table:notation}
\end{table}

\section{Implementation-Aware Partition of ADMA Channel Estimation}\label{sec:pre}
Implementation-aware partition of ADMA channel estimation is first conducted in this section.

\subsection{Setting-Up of ADMA}
For the ease of illustration, we consider a multiuser M-MIMO system, where the BS is equipped with $M$ ($M \gg 1$) antennas in the form of uniform linear array (ULA) and serves $K$ users. We assume that the number of parameters which the BS can handle is $\tau$. In addition, as we presume that each user is equipped with only one antenna, the CM of user-$k$ can be described as a $M \times 1$ vector ${{\mathbf{h}}_k}$. From array signal processing theory, the UL channel vector ${{\mathbf{h}}_k}$ of user-$k$ has the form
\begin{equation}\label{equ:aspt}
{{\mathbf{h}}_k} = {\frac{1}{\sqrt{P}}} {\sum_{p=1}^{P}} {\alpha_{kp}} {\mathbf{a}}({\theta_{kp}}),
\end{equation}
where $P$ is the number of beamforming rays, ${\alpha_{kp}}$ is the complex gain of the $p$-th ray and ${\mathbf{a}}({\theta_{kp}})$ is the array manifold vector which can be expressed as
\begin{equation}
{\mathbf{a}}({\theta_{kp}}) = \left[{1,~e^{j\frac{2{\pi}d}{\lambda}{\sin{\theta_{kp}}}},~...,~e^{j\frac{2{\pi}d}{\lambda}({M-1}){\sin{\theta_{kp}}}}}\right]^T~.
\end{equation}

\begin{Rem}
	In this paper, we do not discuss in the situation that users are equipped with multiple antennas or the propagation signal contains multiple subcarriers in orthogonal frequency duplex division multiplexing (OFDM) systems. In fact, the sparsity of the vectors which can be obtained by collecting the row or column of channel matrices. And so do the sparsity of channel matrices of different subcarriers. So when we obtain the sparsity under ADMA scheme, it can be promoted to plenty of scenarios.
\end{Rem}

\subsection{Channel Sparsity Revealed by ADMA}
To grantee the performance of the proposed channel estimator, the sparsity reveal by ADMA must be kept during the implementation process. The ADMA presents a sparse channel representation for the channel of a M-MIMO system via the Discrete Fourier Transform (DFT) of channel vector, i.e., ${{\mathbf{\tilde h}}_k}$. which can be calculated by
\begin{equation}\label{equ:FFT}
{\mathbf{\tilde h}}_k = {\mathbf{F}}{\mathbf{h}}_k,
\end{equation}
where $\mathbf{F}$ is the $M \times M$ DFT matrix whose element is ${\left[ {\mathbf{F}} \right]_{pq}} = {e^{ - j{\textstyle{\frac{2\pi}{M}}pq}}} / {\sqrt M }$. For the ease of description, there are two lemmas which can be proved from paper \cite{xie2017unified} :

\begin{Lem}
	If $P = 1$ (i.e., AS is zero) and $M \to \infty$, there will be only one non-zero element in ${{\mathbf{\tilde h}}_k}$ and the index of this non-zero element is relative to its DOA or DOD.
\end{Lem}

\begin{IEEEproof}
For $P=1$, ${{\mathbf{h}}_k}$ can be simplified to ${{\mathbf{h}}_k} = {\alpha_{kp}} {\mathbf{a}}({\theta_{kp}})$, then the $b$-th element of ${\mathbf{\tilde h}}_k$ can be calculated as
\begin{equation}\label{equ:lemma_main}
\begin{aligned}
	{\left[ {\mathbf{\tilde h}}_k \right]_{b}} =& {\frac{\alpha_k}{\sqrt{M}}} {\sum_{m=0}^{M-1}} e^{-j({\frac{2\pi}{M}}mb-{\frac{2\pi}{\lambda}md{{\sin}{\theta_{k}}}})}		\\
	=& {\frac{\alpha_k}{\sqrt{M}}} e^{-j({\frac{2\pi}{M}}b-{\frac{2\pi}{\lambda}d{{\sin{\theta_{k}}}}})}		\\
	& \cdot{\frac{{\sin[({\frac{2\pi}{M}}b-{\frac{2\pi}{\lambda}d{{\sin{\theta_{k}}}}})\cdot{\frac{M}{2}}]}}{{\sin[({\frac{2\pi}{M}}b-{\frac{2\pi}{\lambda}d{{\sin{\theta_{k}}}}})\cdot{\frac{1}{2}}]}}},
\end{aligned}
\end{equation}

If $M \to \infty$, we can get that
\begin{equation}\label{equ:lemma_lim}
	\lim_{M \to \infty} \left|{\left[ {\tilde{\mathbf{h}}_k} \right]_{b}}\right| = \left|{\alpha_k}\right|\cdot{\sqrt{M}}\cdot\delta\left({\frac{b}{M}-\frac{d{\sin}{\theta_k}}{\lambda}}\right).
\end{equation}

Eq. \ref{equ:lemma_lim} denotes the relationship between the index of the non-zero element (i.e., $b_0$) in ${\mathbf{\tilde h}}_k$ and the DOA when $M \to \infty$, which can be described as
\begin{equation}\label{equ:reciprocity}
	\left\{
	\begin{aligned}
		b_0 &= \frac{Md\sin{\theta_k}}{\lambda}	\\
		\theta_k &= \arcsin (\frac{b_0\lambda}{Md}),
	\end{aligned}
	\right.
\end{equation}

\end{IEEEproof}

Since we have discussed the situation with $P = 1$ and $M \to \infty$, we can move onto the more complex and realistic scheme:
\begin{itemize}
	\item when $P>1$ and $M \to \infty$, each propagation ray is corresponding to a non-zero element in ${\mathbf{\tilde h}}_k$. The index of the middle element is corresponding to the DOA of user-$k$ while the number of the non-zero elements is corresponding to the AS of user-$k$.
	\item when $P=1$ and $M$ is large but finite, the power leakage emerges because the resolution of the BS is relatively limited, which causes that $b_0 = \frac{Md\sin{\theta_k}}{\lambda}$ is not always an integer. However, there are only a few non-zero elements concentrated around $b_0 = \lfloor \frac{Md\sin{\theta_k}}{\lambda} \rceil$ since $M$ is large. In fact, $M$ denotes the sample precision of the Discrete Time Fourier Transform (DTFT) of ${\mathbf{h}}_k$ in the frequency domain. Since the index of the non-zero elements in ${\mathbf{\tilde h}}_k$ is corresponding to the DOA and AS of user-$k$, $M$ can also determine the spatial resolution of the BS.
	\item when $P>1$ and $M$ is large but finite, it is similar to the situation with $P=1$ and $M$ is large but finite, but the amount of non-zero elements in ${\mathbf{\tilde h}}_k$ will be larger, which is interrelated to the AS of user-$k$.
\end{itemize}

From the above we can see that we can simply get a sparse channel representation by applying DFT to the channel vector and pick the non-zero elements with their indexes. In practical scene, since the BS can handle $\tau$ ($\tau \ll M$) parameters at most, we can use $\tau$ non-zero points of the DFT channel vector ${{\mathbf{\tilde h}}_k}$ instead of all $M$ points to represent the CSI, which can reduce quite a lot of calculating and feedback overhead.

\subsection{Sparsity Enhancer for ADMA}
To enhance the channel sparsity under ADMA scheme, we define:
\begin{Def}
	Define ${\mathbf{\Phi }}(\phi_k )$ as the rotation matrix for user-$k$ which can be expressed as
	\begin{equation}
		{\mathbf{\Phi }}(\phi_k ) = {\rm{diag}}\left\{ {\left[ {1,{e^{j\phi_k }}, \ldots ,{e^{j(M - 1)\phi_k }}} \right]} \right\},
	\end{equation}
	where $\phi_k \in \left[ { \text{-} {\textstyle{\frac{\pi}{M}}},{\textstyle{\frac{\pi}{M}}}} \right]$.
	Then we can add this rotate-operation to the DFT calculating. Define ${{\mathbf{\tilde h}}^{{\rm{ro}}}}_k$ as the new channel representation with rotation given by
	\begin{equation}
		{{\mathbf{\tilde h}}^{{\rm{ro}}}}_k = {\mathbf{F}{\Phi}}(\phi_k ){\mathbf{h}}_k.
	\end{equation}
\end{Def}
In this way, we can use less non-zero elements to represent the channel vector. Or in practical scene, the $\tau$ non-zero elements we pick will contain more energy of the channel, which is a great benefit for the training overhead.
\begin{Rem}
	Notice that the rotation is actually the translation of ${{\mathbf{\tilde h}}_k}$ in the frequency domain. Since the spatial resolution of the BS is relatively limited, we can get the sample points aligned with the middle of the peak of the DTFT of ${\mathbf{h}}_k$ to the greatest extent via the rotation operation. Since the sampling interval in the frequency domain is ${\frac{\pi}{M}}$, it is only necessary to search over $\phi_k \in \left[ {\text{-} {{{\frac{\pi}{M}}},{\frac{\pi}{M}}}} \right]$
\end{Rem}
In this case, we can define the index set to describe the signature of channel vectors with rotation as following:
\begin{Def}
	Define ${\cal B}^{\rm{ro}}_k$ as the spatial signature of user-$k$ which can be determined according to
	\begin{equation}\label{twopara}
		\max_{{\phi_k},{\cal B}^{\text{ro}}_k} \left.\left|\left| {\left[{\tilde{{\mathbf{h}}}}^{\text{ro}}_k\right]}_{{\cal B}^{\text{ro}}_k} \right|\right|^2 \over \left|\left|{\tilde{{\mathbf{h}}}}^{\text{ro}}_k \right|\right|^2 \right.~,~~{\text{subject to}}~\left| {{\cal B}^{\text{ro}}_k} \right|=\tau,
	\end{equation}
\end{Def}
Now we have two parameters for each user to be determined under ADMA scheme: $\phi$ and ${\cal B}^{\rm{ro}}$. The main benefit of this sparse channel representation is that we only need a few training sequences because users from different directions whose DOAs do not overlap can share the same training sequence. In practical scene, we usually use $\tau$ orthogonal training sequence which can make full use of the BS.

Meanwhile, we can explain the transmission strategy under ADMA scheme which can be divided into three stages: the preamble stage, the UL training stage and the DL training stage. The aim of the preamble stage is to collect the two parameters of all users and divide them into different groups according to their spatial signatures. Then in the UL training stage and the DL training stage we can perform faster estimation than conventional channel estimation methods due to the grouping in the preamble stage. The preamble stage is not necessary after each UL and DL training stage and the times for UL and DL training stages after one preamble stage is corresponding to the mobility of users.

\subsection{Preamble Module}
In the preamble period, we need to find $\phi$ and ${\cal B}^{\rm{ro}}$ for each user so that we can allocate all users into different groups in which the index sets, i.e., ${\cal B}^{\rm{ro}}$ of users do not overlap each other's.

First we allocate all $K$ users into $G$ groups, each containing $\tau$ users as the BS can handle up to $\tau$ training sequences. Then we apply the conventional UL training for each group, and the receiving signals matrix of each group in the BS is given by
\begin{equation}
{\mathbf{Y}} = {\mathbf{H}}{{\mathbf{D}}^{{\raise0.5ex\hbox{$\scriptstyle 1$}
			\kern-0.1em/\kern-0.15em
			\lower0.25ex\hbox{$\scriptstyle 2$}}}}{{\mathbf{S}}^H} + {\mathbf{N}} = \sum\limits_{i = 1}^\tau  {\sqrt {{d_i}} {{\mathbf{h}}_i}{\mathbf{s}}_i^H}  + {\mathbf{N}},
\end{equation}
where ${\mathbf{Y}}\in{\mathbb{C}^{M \times L}}$, ${\mathbf{H}}=\left[ {{{\mathbf{h}}_1}, \ldots ,{{\mathbf{h}}_\tau }} \right] \in {\mathbb{C}^{M \times \tau}}$,
${\mathbf{S}} = \left[ {{{\mathbf{s}}_1}, \ldots ,{{\mathbf{s}}_\tau }} \right] \in {\mathbb{C}^{L \times \tau }}$,
${\mathbf{D}} = {\rm{diag}}\left\{\left[ {{{{d}}_1}, \ldots ,{{{d}}_\tau }} \right]\right\}\in {\mathbb{C}^{\tau  \times \tau }}$ and ${d_k}=\left.{P_k^{ut}}/{L\sigma _p^2}\right.$
is used to satisfy the energy constraint (${P_k^{\rm{ut}}}$ is the UL training energy constraint of user-$k$, and $\sigma _p^2$ is the pilot
signal training power),
${\mathbf{N}}\in{\mathbb{C}^{{M} \times L}}$ is the additive white Gaussian noise matrix. Then ${{\mathbf{h}}_k}$ can be calculated through linear square (LS) method as

\begin{equation}\label{equ:preamble_2}
{{\mathbf{\hat h}}_k} = \frac{1}{{\sqrt {{d_k}} L\sigma _p^2}}{\mathbf{Y}}{{\mathbf{s}}_k}.
\end{equation}

Then we can find $\phi_k$ and ${{\cal B}_k^{\rm{ro}}}$ for each user by adopting Eq. (\ref{twopara}). The specific method is discussed in Section \ref{sec:sim}. After that, we need to allocate all users into $G^{\rm{ul}}$ groups in which the index sets of users do not overlap each other's so that the users in the same group can share the same training consequence, which can be described as
\begin{equation}\label{equ:grouping}
	\left\{
	\begin{aligned}
		&{\cal B}^{\rm{ro}}_k \cap {\cal B}^{\rm{ro}}_l	= \emptyset\\
		&\min\left|b_1-b_2\right| \geq \Omega~,~\forall b_1 \in {\cal B}^{\rm{ro}}_k~,~\forall b_2 \in {\cal B}^{\rm{ro}}_l,
	\end{aligned}
	\right.
\end{equation}
where $\Omega$ is a certain guard interval which depends on the tolerance of users for the interference due to pilot reusing. Here we present a grouping strategy that is easy for VLSI implementation in Section \ref{sec:hard}.

\subsection{UL Training Module}
In the UL training, all $K$ users send their training sequences to the BS. The received signals matrix in the BS is given by
\begin{equation}
{\mathbf{Y}} = \sum\limits_{i = 1}^{{G^{{\rm{ul}}}}} {\sum\limits_{k \in {{\cal U}}_i^{{\rm{ul}}}} {\sqrt {{d_i}} } } {{\mathbf{h}}_k}{\mathbf{s}}_i^H + {\mathbf{N}}.
\end{equation}

So first we extract the channel vector for group-$g$ through a conventional LS method:
\begin{equation}
	{\mathbf{y}}_g = {\frac{1}{{L\sigma_p^2 }}}{\mathbf{Y}}{{\mathbf{s}}_g}.
\end{equation}

Since the two parameters of each user is different, we should extract ${{\mathbf{\tilde h}}_k}$ for each user-$k$ through
\begin{equation}\label{equ:UL_1}
	{\widehat {\left[ {{{{\mathbf{\tilde h}}}^{{\rm{ro}}}}_k} \right]}_{{{\cal B}}_k^{{\rm{ro}}}}} = {\left[ {{\mathbf{\tilde y}}_{g,k}^{{\rm{ro}}}} \right]_{{{\cal B}}_k^{{\rm{ro}}}}} = {\left[ {\frac{1}{{\sqrt {{d_k}} }}{\mathbf{F\Phi }}({\phi _k}){{\mathbf{y}}_g}}\right]_{{{\cal B}}_k^{{\rm{ro}}}}}.
\end{equation}

Finally we can recover the ${\mathbf{\hat h}}_k$ for user-$k$ by
\begin{equation}\label{equ:UL_2}
{{\mathbf{\hat h}}_k} = \Phi {({\phi _k})^H}{{\mathbf{F}}^H}{\mathbf{\hat {\tilde h}^{{\rm{ro}}}_k}} = \Phi {({\phi _k})^H}{\left[ {{{\mathbf{F}}^H}} \right]_{:,{{\cal B}}_k^{{\rm{ro}}}}}{\widehat {\left[ {{{{\mathbf{\tilde h}}}^{{\rm{ro}}}}_k} \right]}_{{{\cal B}}_k^{{\rm{ro}}}}}.
\end{equation}

\subsection{DL Training Module and Its Reciprocity}
Based on the reciprocity of ADMA, the DL CSI can be easily obtained from the UL training as shown in \cite{xie2017unified}. The reciprocity of ADMA comes from that the propagation path of electromagnetic wave is reciprocal. As a result, the DOA (DOD) of DL signal is the same as the DOD (DOA) of the UL signal. Assume that the DL spatial signature of user-$k$ is $\overline{{\cal B}^{\rm{ro}}_k}$ which can determined by the UL spatial signature ${{\cal B}^{\rm{ro}}_k}$ by applying Eq. (\ref{equ:reciprocity}):
\begin{equation}
	\sin{\theta_{kp}}=\frac{q\lambda^{\rm{ul}}}{Md}=\frac{\overline{q}\lambda^{\rm{dl}}}{Md},
\end{equation}
where $q$ and $\overline{q}$ are the elements in ${{\cal B}^{\rm{ro}}_k}$ and $\overline{{\cal B}^{\rm{ro}}_k}$, while $\lambda^{\rm{ul}}$ and $\lambda^{\rm{dl}}$ denote the UL and DL carrier wavelengths. Since $\sin{\theta_{kp}}$ is a monotonic function with $\theta_{kp} \in \left[ \textstyle{-}\frac{\pi}{2} , \frac{\pi}{2} \right]$,
the minimum and maximum elements of ${{\cal B}^{\rm{ro}}_k}$ and $\overline{{\cal B}^{\rm{ro}}_k}$ have an one-to-one correspondence, i.e.,:
\begin{equation}
	\overline{q}_{\rm{min}}=\biggl\lfloor {\frac{\lambda^{\rm{ul}}}{\lambda^{\rm{dl}}}q_{min}}\biggr\rfloor,~~~\overline{q}_{\rm{max}}=\biggl\lceil {\frac{\lambda^{\rm{ul}}}{\lambda^{\rm{dl}}}q_{max}}\biggr\rceil,
\end{equation}
where $q_{\rm{min}} \leq q \leq q_{\rm{max}},~\forall q \in {{\cal B}^{\rm{ro}}_k}$. Meanwhile, $\overline{\phi_k}$ can be calculated by ${\overline{\phi}}_k = (\lambda^{\rm{ul}}/\lambda^{\rm{dl}}) \phi_k$ similarly.

The DL training is mostly the same with UL training except the Grouping strategy. In DL training module, since users with identical spatial signatures can be carried out with the same beamforming vectors simultaneously, they can share the same training sequence. Meanwhile, users whose spatial signatures do not overlap each other's can share the same training sequence, which is the same with the UL training Grouping strategy. As a result, we denote our DL training strategy in two steps. First we allocate users with identical spatial signatures into the same cluster. Then we allocate these clusters in to different groups through Eq. (\ref{equ:grouping}). The rest of transmission and estimation is the same with the UL training module.

With the successful algorithm partition, we are now able to carry out the detailed implementation-aware algorithm optimization and module-wise architecture design as follows.

\section{Approximation and Quantization}\label{sec:sim}
For simulations, the mean square error (MSE) is calculated as follows:
\begin{equation}
{\text{MSE}} = \frac{{\mathbb{E}}\{{|| {\mathbf{h}}_k - {\mathbf{\hat h}}_k|| }^2\}}{{\mathbb{E}}\{{|| {\mathbf{h}}_k || }^2\}}.
\end{equation}
For comparison, the system parameters are set as: $M=128,~K=32,~L=64,~\tau=16,~{\theta_k} \in \{-48.59^{\circ}, -14.48^{\circ}, 48.59^{\circ}, 14.48^{\circ}\}~and~\Delta \theta_k = 2^{\circ}$, which are consistent with those in \cite{xie2017unified}.

\subsection{Approximation for Sliding Window Method}
\begin{figure}[htbp]
	\centering
	\includegraphics[width = .8\linewidth]{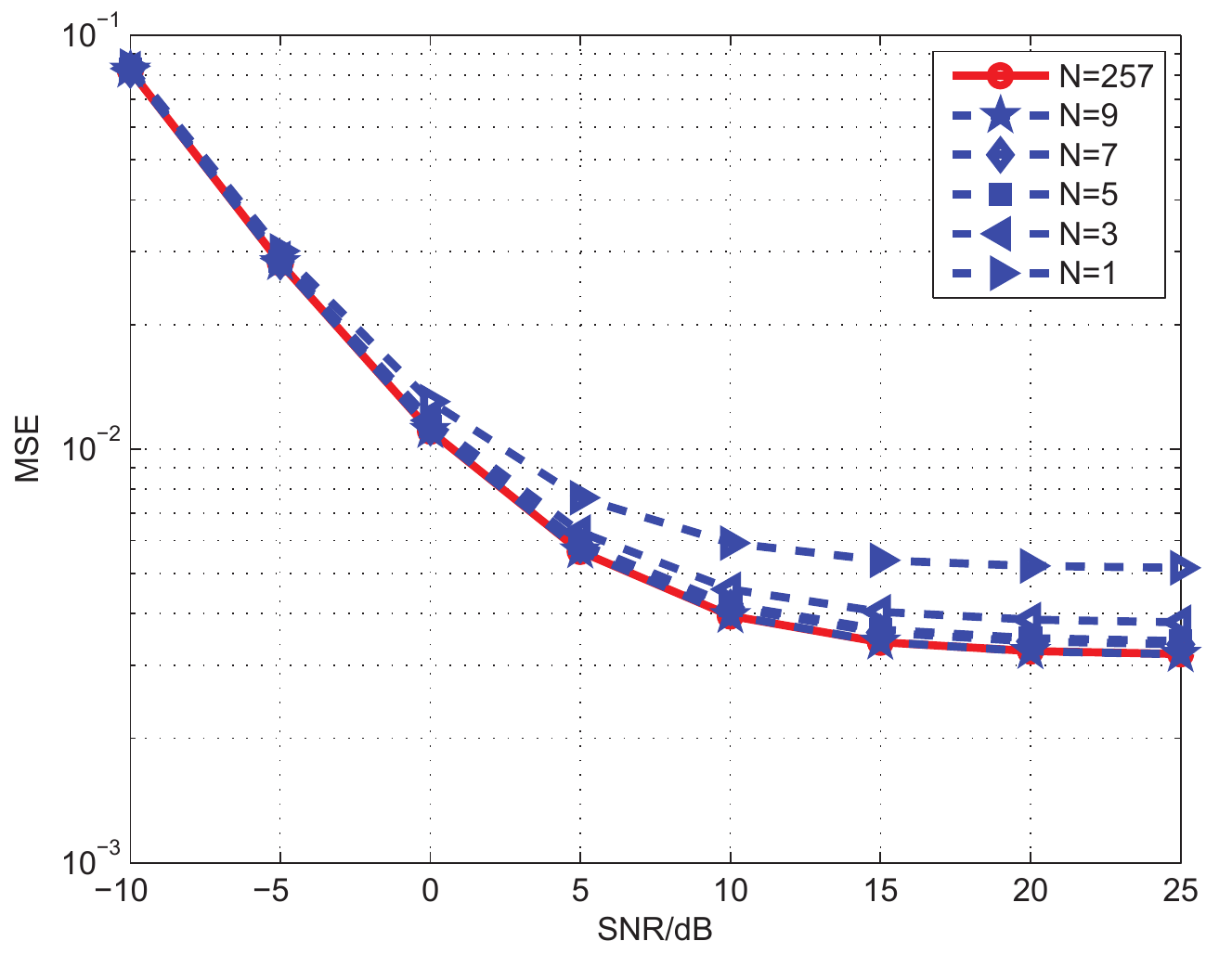}
	\caption{MSE results of different $N$.}
	\label{fig:epsN}
\end{figure}

The authors of \cite{xie2017unified} proposed a basic way to find ${{\cal B}_k^{\rm{ro}}}$ for user-$k$ by adopting a one dimensional search over $\phi_k = \left[ { \text{-} {\textstyle{\frac{\pi}{M}}},{\textstyle{\frac{\pi}{M}}}} \right]$ and for each possible $\phi_k$
by sliding a window of size $\tau$ over the $M$ elements in ${\mathbf{\tilde h}_k}$ to determine ${{\cal B}_k^{\rm{ro}}}$ that maximizes the channel power ratio. However, there are two main problems if we operate through this method. The first one is that searching over $\phi_k  = \left[ { \text{-} {\textstyle{\frac{\pi}{M}}},{\textstyle{\frac{\pi}{M}}}} \right]$ can not be carried out since it is a continuous interval.
Fig. \ref{fig:epsN} shows the corresponding MSE of the simulations with $N$ separate elements we choose in the ${\left[ {\text{-} {\frac{\pi}{M}}, {\frac{\pi}{M}}} \right]}$. As we can see from that, the MSE of $N = 3$ is nearly the same as those of $N>3$, so $N = 3$ turns out to be suitable for VLSI implementation.

The second problem is that this method will introduce quite a lot latency and increase the computation complexity as the accumulator and divider are needed. Here we have some approximations to lower the complexity:

\begin{itemize}
	\item The first one is to find the maximum element in $| {{\mathbf{\tilde h}}^{{\rm{ro}}}}_k |$ for each possible $\phi_k$ and determine the best ${b_k^{\rm{ro}}}$ and $\phi_k$ for user-$k$ from the largest elements of all the possible $\phi_k$.
	\item The second one is to find the maximum element and the second largest element in $| {{\mathbf{\tilde h}}^{{\rm{ro}}}}_k |$ and calculate the quadratic sum of the largest two elements for each possible $\phi_k$.  Determine the best ${b_k^{\rm{ro}}}$ and $\phi_k$ from the largest quadratic sum of all the possible $\phi_k$ (the index ${b_k^{\rm{ro}}}$ will be the mean value of the indexes of the largest two elements in $| {{\mathbf{\tilde h}}^{{\rm{ro}}}}_k |$ with $\phi_k$) .
	\item The third one is to find the maximum element in $| {{\mathbf{\tilde h}}^{{\rm{ro}}}}_k |$ and calculate the quadratic sum of $\tau$ continuous elements which center on the maximum element for each possible $\phi_k$. Determine the best ${b_k^{\rm{ro}}}$ and $\phi_k$ from the largest quadratic sum of all the possible $\phi_k$ (the index ${b_k^{\rm{ro}}}$ comes from the largest elements in $| {{\mathbf{\tilde h}}^{{\rm{ro}}}}_k |$ with $\phi_k$) .
\end{itemize}

\begin{figure}[htbp]
	\centering
	\includegraphics[width = 0.8\linewidth]{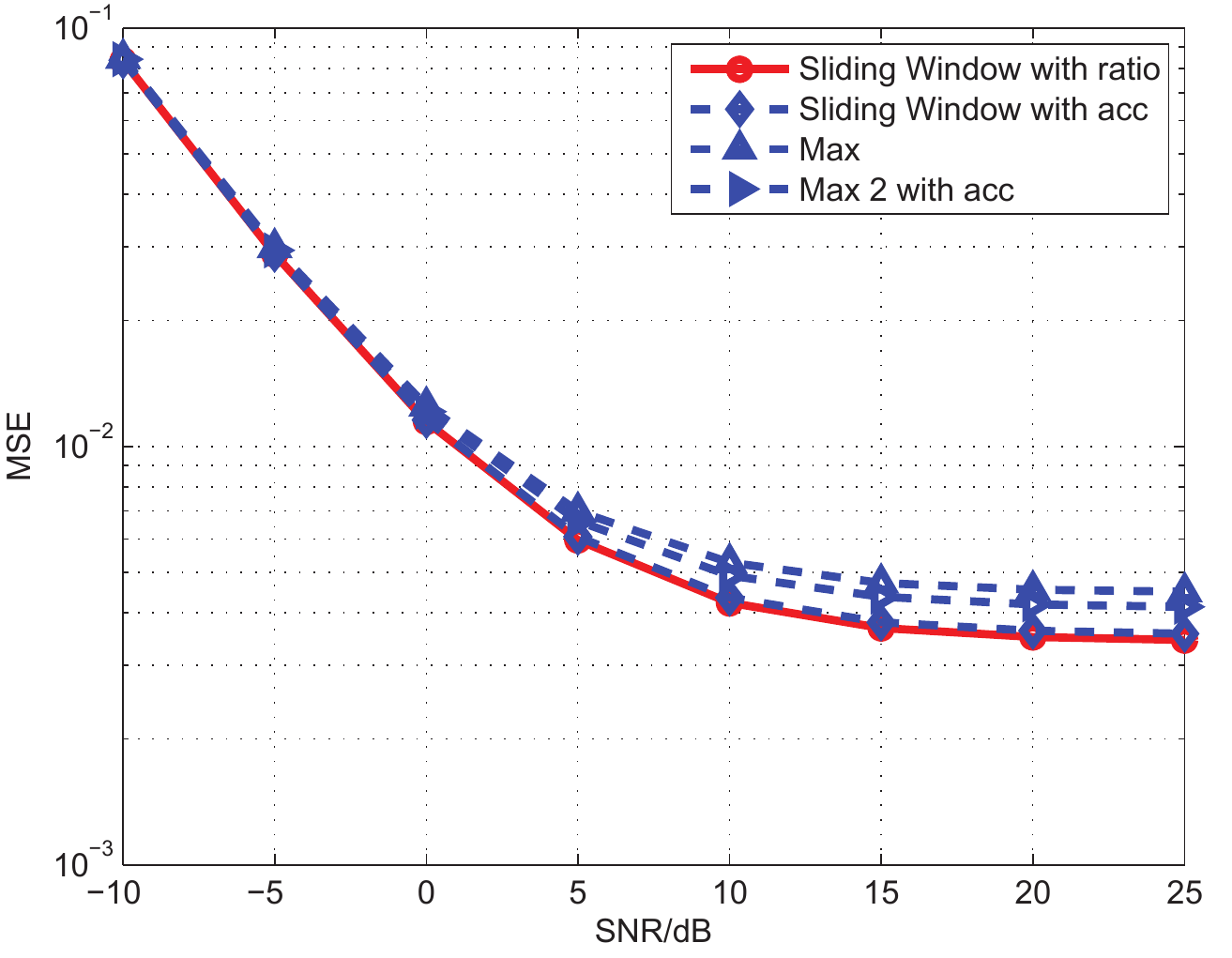}
	\caption{MSE of different methods to determine the spatial signature of user.}
	\label{fig:epsswvsmax}
\end{figure}

Fig. \ref{fig:epsswvsmax} shows the MSE increase for the approximations above. It is obvious that the performance of the third approximation is nearly the same as the basic method with a divider economized. Meanwhile, the performance loss of the first or the second method is higher but relatively acceptable with only one or two comparators needed.

\subsection{Quantization Scheme}
For quantization, the variables are quantified with $1$ sign bit, $p$ integral bits, and $q$ fractional bits which is expressed as fixed [$1,p,q$]. The width of integer $p$ is usually determined by the Probability Density Function (PDF) of the data. But in our algorithm the largest data must be less than $2^q$ since the channel state information contains the largest element in $| {{\mathbf{\tilde h}}^{{\rm{ro}}}}_k |$. Here we show the statistics of large amount of the largest data in ${\mathbf{h}}_k$, ${{\mathbf{\tilde h}}^{{\rm{ro}}}}_k$ and $| {{\mathbf{\tilde h}}^{{\rm{ro}}}}_k |$.
According to our statistics, the largest data is less than $2^8$ so that the width of integral part of variables is set as $8$.

\begin{figure}[htbp]
	\centering
	\subfigure[The statistics of large amount of the largest data in ${\mathbf{h}}_k$.]{
		\includegraphics[width = 0.46\linewidth]{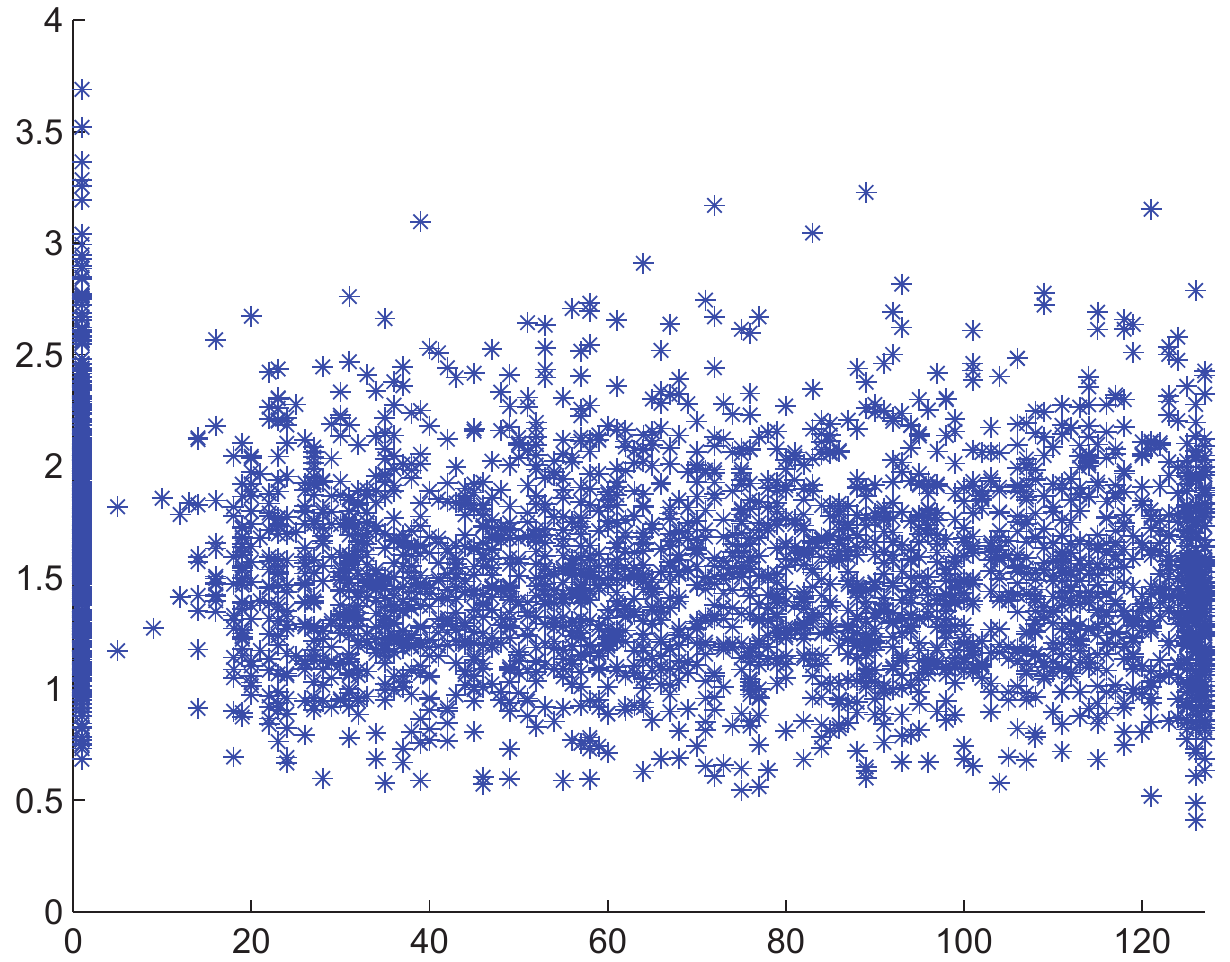}
	}
	\subfigure[The statistics of large amount of the largest data in ${{\mathbf{\tilde h}}^{{\rm{ro}}}}_k$.]{
		\label{fig:epsfix1}
		\includegraphics[width = 0.46\linewidth]{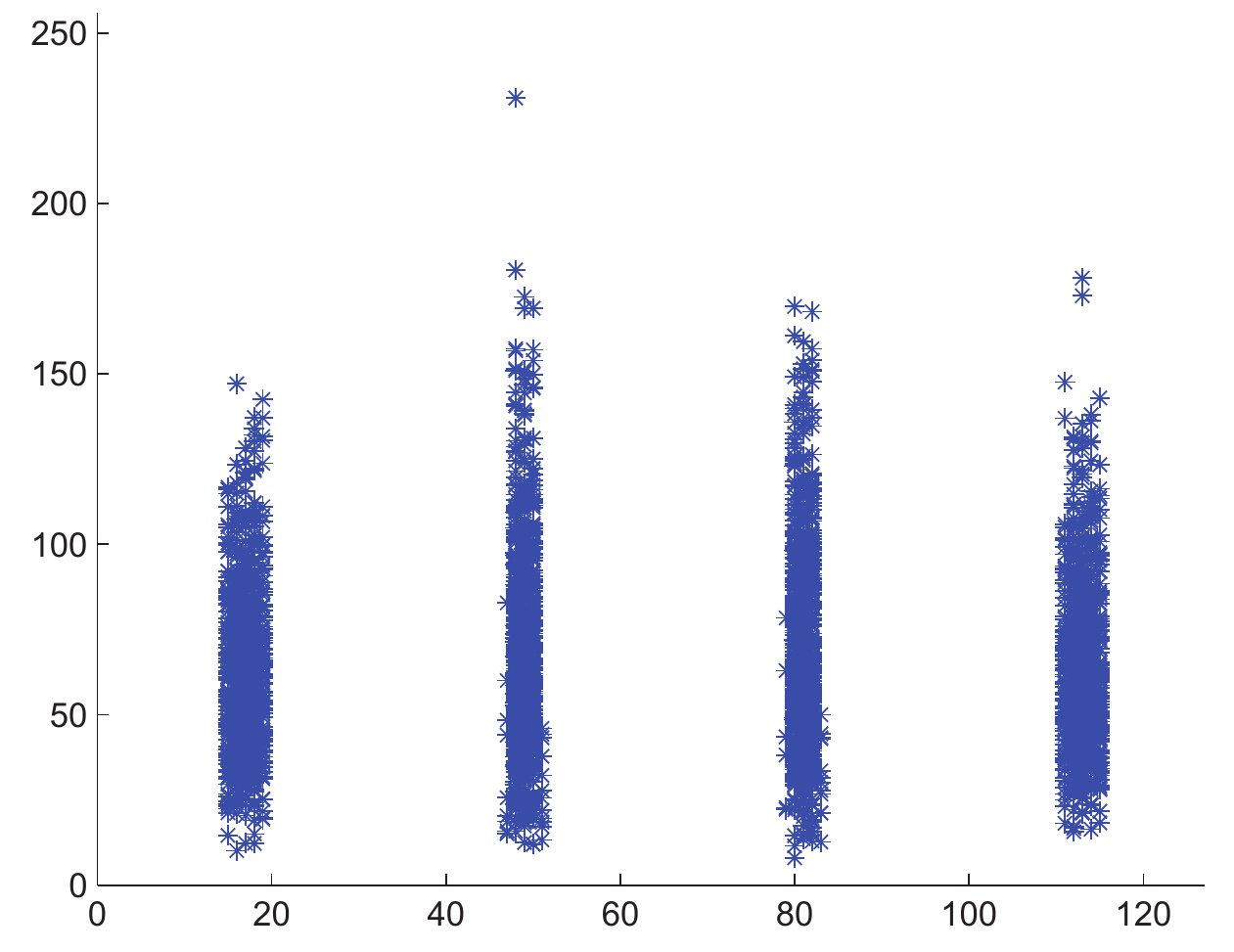}
	}
	\subfigure[The statistics of large amount of the largest data in $\left| {{\mathbf{\tilde h}}^{{\rm{ro}}}}_k \right|$.]{
		\label{fig:epsfix2}
		\includegraphics[width = .8\linewidth]{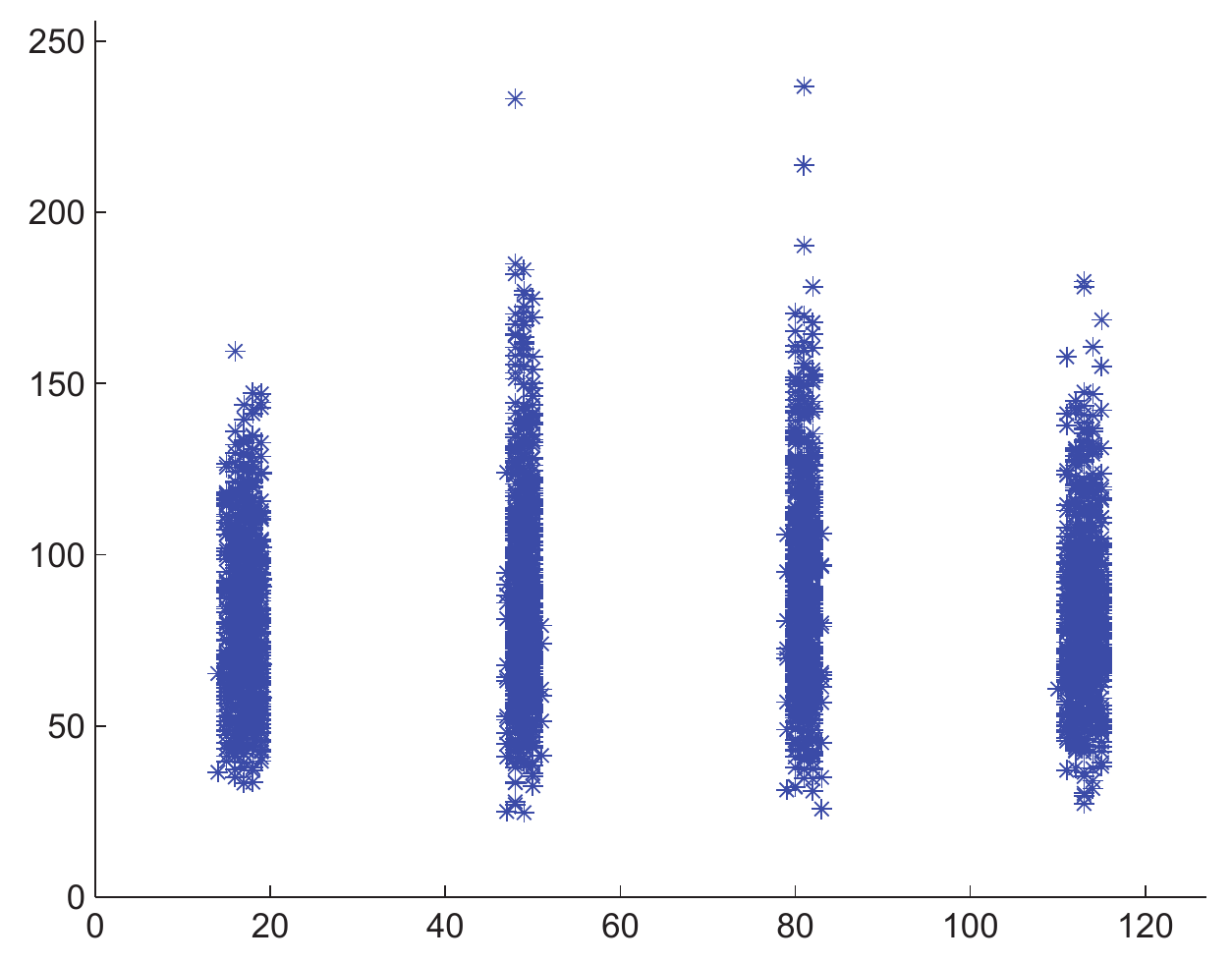}
	}
	\caption{The statistics of large amount of the largest data in ${\mathbf{h}}_k$, ${{\mathbf{\tilde h}}^{{\rm{ro}}}}_k$ and $| {{\mathbf{\tilde h}}^{{\rm{ro}}}}_k |$. The maximum data appears in the element of $| {{\mathbf{\tilde h}}^{{\rm{ro}}}}_k |$. }
	\label{fig:epsfix:old}
\end{figure}

\begin{figure}[htbp]
	\centering
	\includegraphics[width = .8\linewidth]{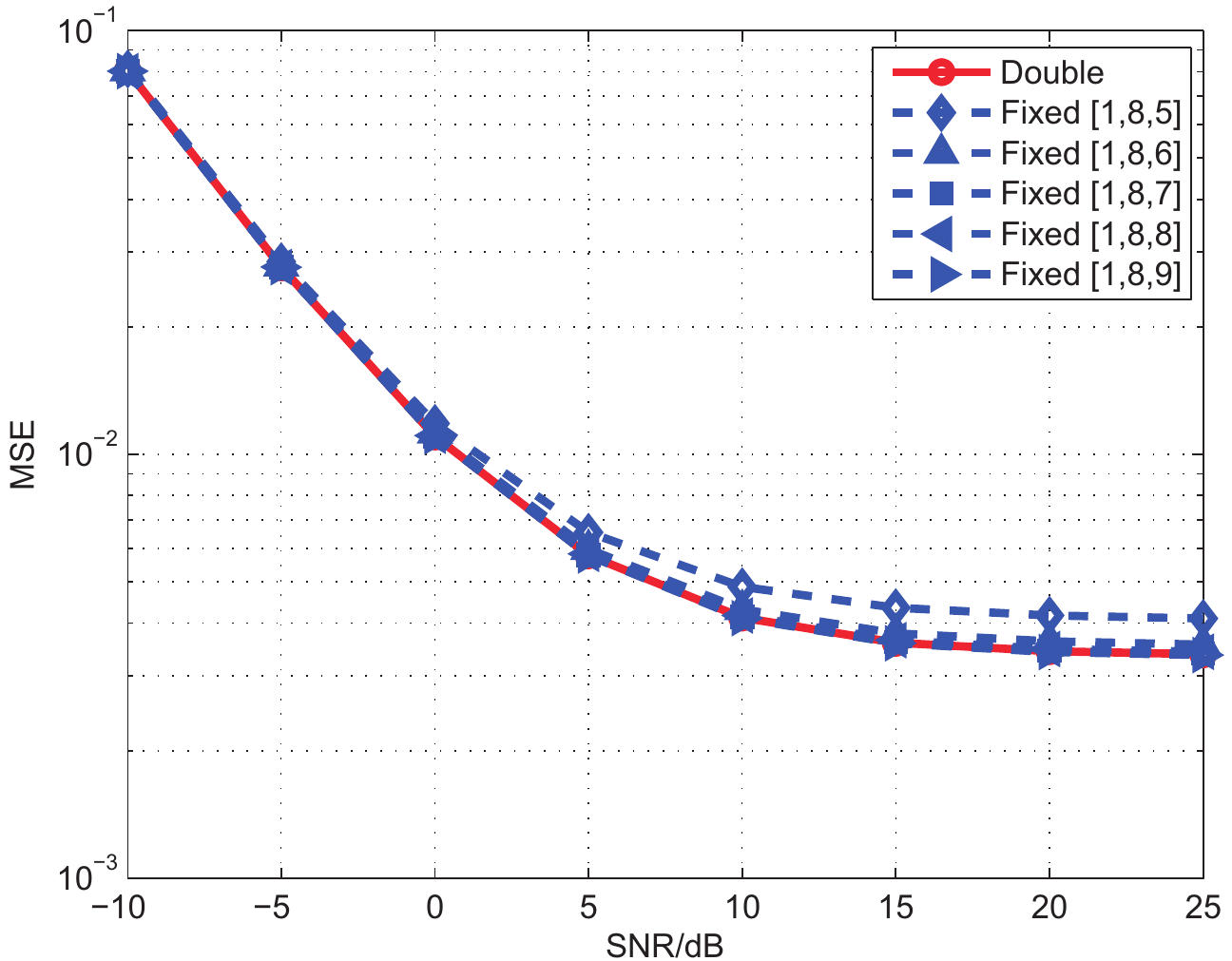}
	\caption{MSE results of double precision floating and fixed simulations.}
	\label{fig:epsfix}
\end{figure}

\begin{figure*}[htbp]
	\centering
	\includegraphics[width = \linewidth]{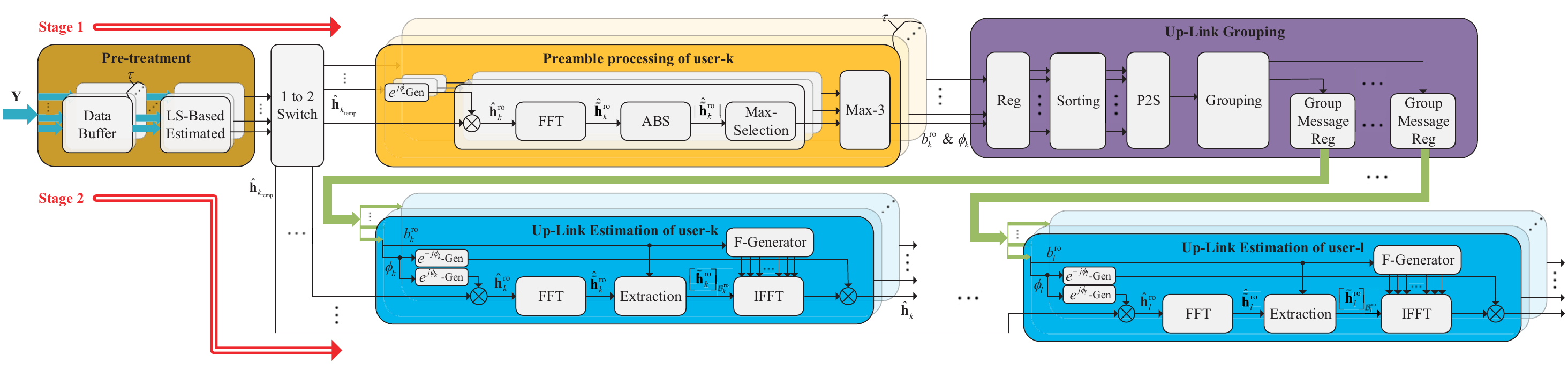}
	\caption{The overall hardware architecture of channel estimation under ADMA scheme. The number of preamble processing module is equal to the number of training sequences $\tau$ and the number of UL estimation module is equal to the number of users in order to achieve the highest processing speed. } \label{fig:pdftop}
\end{figure*}

\begin{figure*}[htbp]
	\centering
	\includegraphics[width = .9\linewidth]{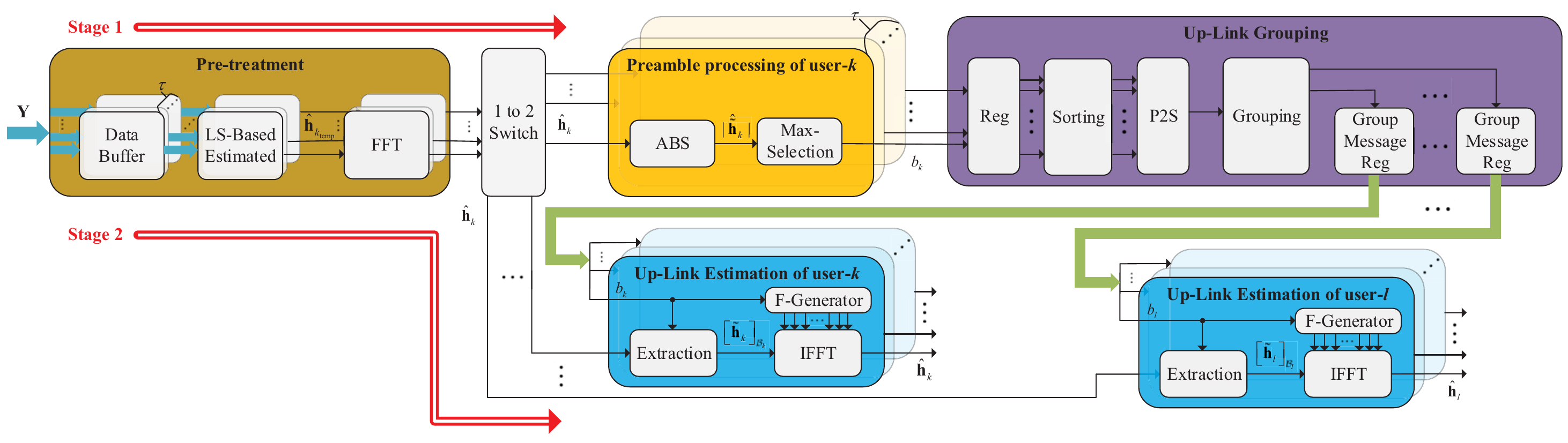}
	\caption{The overall hardware architecture without rotation operations. } \label{fig:pdftopwithoutro}
\end{figure*}

In order to determine the width of fractional part of the variables, the corresponding MSE of double floating and fixed simulations are illustrated in Fig. \ref{fig:epsfix}. From Fig. \ref{fig:epsfix} the MSE of fixed [$1,8,6$] and fixed [$1,8,7$] simulation keep almost the same, with a slight degradation compared with the double floating simulation. However, the MSE of fixed [$1,8,5$] simulation is a relatively far from double floating simulation. As a result, the quantization scheme with fixed [$1,8,6$] may be preferred for hardware implementation.

\section{Pipelined Architecture}\label{sec:hard}

For channel estimation under ADMA scheme, the operations are conducted on the $M$-dimensional vectors and matrices, where $M$ is large. For low-complexity and high processing speed, the pipelined architecture is demonstrated in Fig. \ref{fig:pdftop}. In addition, the quantization scheme with ``fixed [$1,8,6$]'' is employed, together with $\phi_k = {\{ \text{-} {\textstyle{\frac{\pi}{M}}}, 0, {\textstyle{\frac{\pi}{M}}}\}}$, respectively.

Our design has two stages controlled by a 1-to-2 switch. Stage 1 consists of pre-treatment module, preamble processing module and UL-grouping module corresponding to Eq.s (\ref{equ:preamble_2}) and (\ref{equ:FFT}). Stage 2 comprises pre-treatment module and UL-Estimation module corresponding to Eq.s (\ref{equ:UL_1}) and (\ref{equ:UL_2}).

\subsection{Module Design}
\subsubsection{Pre-treatment Module}
\begin{figure}[htbp]
	\centering
	\includegraphics[width = .95\linewidth]{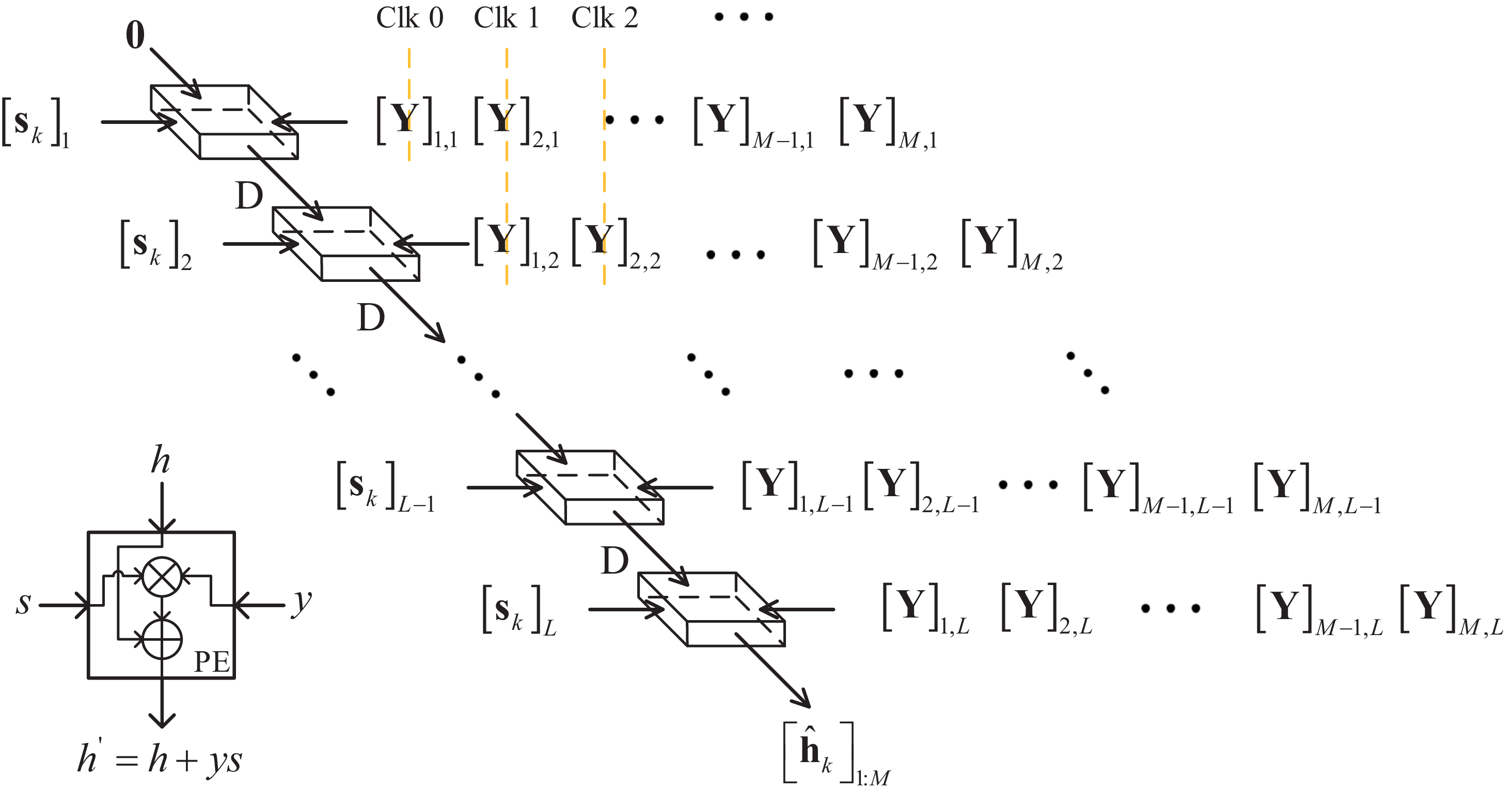}
	\caption{Systolic structure of LS-based estimation module. }
	\label{fig:systolic}
\end{figure}

The pre-treatment module can be reused since preamble module and UL-estimation module are processed in different time slots. Pre-treatment module consists of data buffer and LS-based estimation module. The LS-based estimation module corresponding to Eq. (\ref{equ:preamble_2}) can be implemented by a systolic structure \cite{urquhart1984systolic} whose data flow graph is shown in Fig. \ref{fig:systolic}, which is an efficient processing method for matrix-vector multiplication. The processing element (PE) performs one complex multiplication and one complex addition. Each PE is corresponding to one elements of training sequence ${\mathbf{s}}_k$ and one column of receiving data matrix ${\mathbf{Y}}$ so that the data buffer is needed to get the data transmission proper because ${\mathbf{Y}}$ is received by column (i.e., we receive $M$ elements in one column of ${\mathbf{Y}}$ in one clock period).

\subsubsection{Fast Fourier Transform (FFT) Module}
\begin{figure}[htbp]
	\centering
	\includegraphics[width = \linewidth]{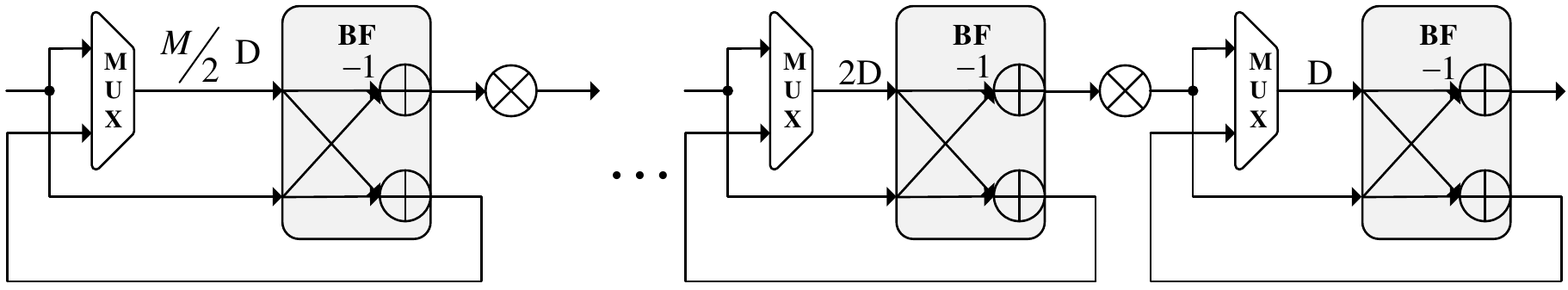}
	\caption{Feed-back pipelined hardware architecture of FFT module.}
	\label{fig:fft}
\end{figure}

Eq. (\ref{equ:FFT}) can be divided into two steps. One is a diagonal matrix and vector multiplication which can be implemented by a complex multiplier and a $\Phi$-generator which outputs the diagonal elements of ${\Phi}(\phi_k)$ in pipeline. The other is DFT which can be implemented by Fast Fourier Transform (FFT) processors, reducing the computational complexity to ${\cal} O(M{\text{log}_2}M)$. These are plenty of structures of FFT which emphasize either higher processing speed or less resources overhead \cite{ayinala2012pipelined,cheng2007high,chang2008efficient}. For higher hardware efficiency, the single-path feedback pipelined hardware architecture is employed as it is shown in Fig. \ref{fig:fft}, where the number of registers is the smallest as a result of the application of multiplexers.

\subsubsection{Up-link Grouping module}

\begin{figure*}[htbp]
	\centering
	\includegraphics[width = .95\linewidth]{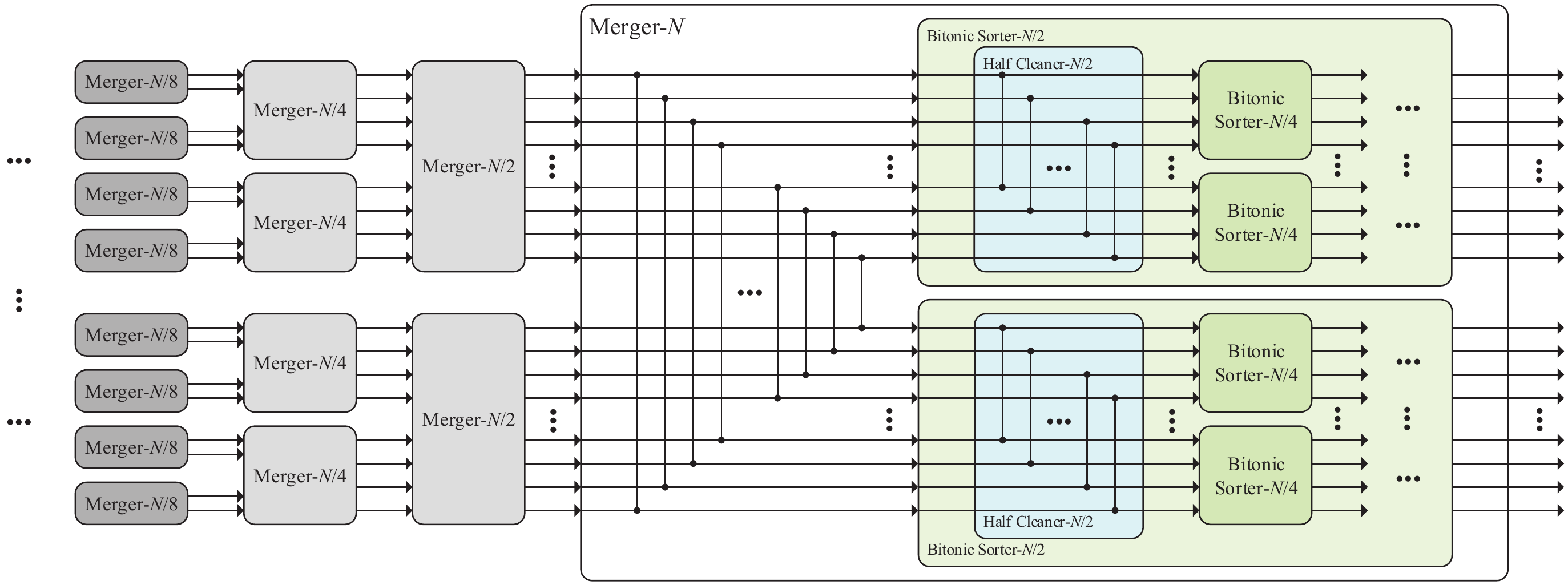}
	\caption{Merging network structure of $N$-element sorting.}
	\label{fig:mergingnetwork}
\end{figure*}

\begin{figure}[htbp]
	\centering
	\subfigure[Systolic structure of Grouping module.]{

		\includegraphics[width = \linewidth]{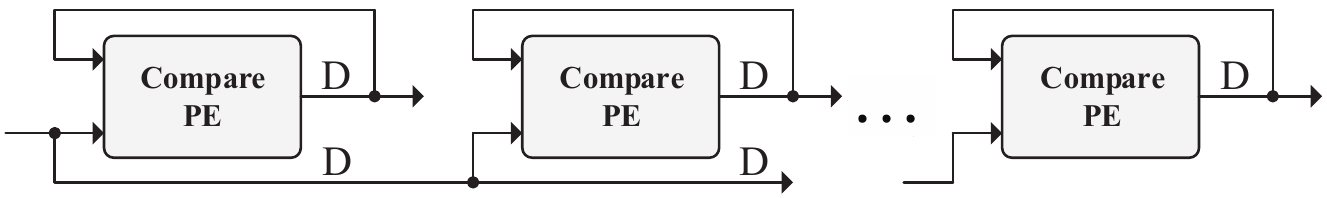}
	}
	\subfigure[The structure of Compare PE.]{
		\label{fig:Groupingpe}
		\includegraphics[width = .65\linewidth]{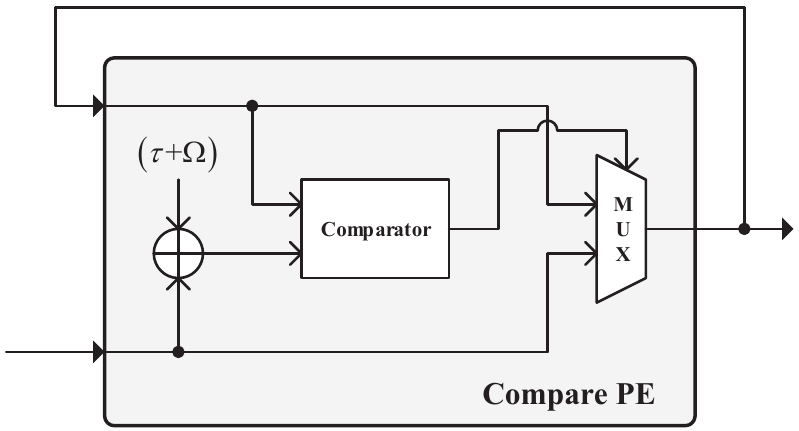}
	}
	\caption{Systolic structure of Grouping module.}
	\label{fig:Grouping}
\end{figure}

In the Up-link Grouping module, there are two main submodules: sorting and grouping. The sorting module is implemented by merging network \cite{batcher1968sorting} in pipeline which is shown in Fig. \ref{fig:mergingnetwork}. This sorting network is mainly based on recursion, merging from 2-element comparison to $N$-element comparison (assuming $log_2 N$ is a positive integer). Meanwhile, the merger-$N$ module is a combination of symmetric comparing network and two bitonic sorter of $N/2$ elements. Then the bitonic sorter-$N/2$ can be implemented by a half cleaner-$N/2$ module and two bitonic sorter-$N/4$. The grouping module is implemented by a systolic structure shown in Fig. \ref{fig:Grouping} where each comparing PE is corresponding to a group and decide if the latest input ${b_k^{\rm{ro}}}$ is suitable for the group by comparing it with the latest ${b_l^{\rm{ro}}}$ in this group.
Since the outputs of sorting module is paralleled and the input of grouping module is serial, a parallel-to-serial module is necessary.

\begin{Rem}
	Notice that the grouping messages are sent to users through a independent feedback channel which is not contained in our hardware design.
\end{Rem}

\subsubsection{Up-link estimation module}
In the Up-link estimation module, the implementation of Eq. (\ref{equ:UL_1}) is a combination of a complex multiplier, an FFT module and an extraction module. Besides, the implementation of Eq. (\ref{equ:UL_2}) consists of an Inverse Fast Fourier Transform (IFFT) module and a complex multiplier. Due to the sparsity of ${\mathbf{\tilde h}}_k$, the IFFT module can be treated as an $M \times \tau $ matrix and $\tau \times 1$ vector multiplication which can be implemented by systolic structure which consists of $\tau$ PEs for higher efficiency.

\subsection{Optimized Architecture without Rotation}
As we can see from the Fig. \ref{fig:pdftop}, the FFT modules of preamble processing module and Up-link estimation module could be reused since they are not deployed at the same time. However, the spatial signatures of users in the same group are different, leading to the waste of FFT module. Here we find that we can simply omit the rotation operations as the architecture shown in Fig. \ref{fig:pdftopwithoutro}, which reuses the FFT module and reduces the number of FFT modules from $\tau + K$ to $\tau$, saving the resources a lot.
\begin{figure*}[htbp]
	\centering
	\includegraphics[width = \linewidth]{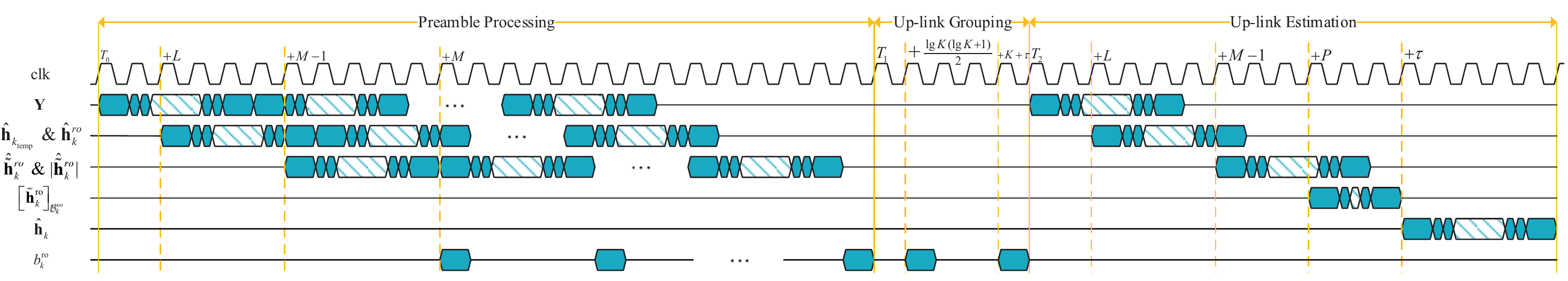}
	\caption{The processing schedule for the system.} \label{fig:timeoptimistic}
\end{figure*}
\begin{table*}[htbp]
	\centering
	\caption{Resource Cost of the Proposed Estimator}
	\begin{tabular}{ l  l  l  l  l }
        \Xhline{1.0pt}
		Modules & Complex Multipliers & Complex Adders & Real Comparators & Registers \\
		\hline
		\rowcolor{mygray}
		LS-based Estimation & $L$ & $L$ & 0 & $L-1$ \\
		FFT & ${\log _{_2}}M-1$ & $2{\log _{_2}}M$ & 0 & $M-1$ \\
		\rowcolor{mygray}
		ABS & 1 & 0 & 0 & 0 \\
		Max-selection & 0 & 0 & 1 & 1 \\
		\rowcolor{mygray}
		Sorting & 0 & 0 & $K{\log _2}K$ & $K{\log_2 K(\log_2 K + 1)}/{2}$ \\
		Grouping & 0 & 1 & $\tau$ & $2\tau$ \\
		\rowcolor{mygray}
		Extraction & 0 & 0 & 1 & 0 \\
		IFFT & $\tau$ & $\tau$ & 0 & $\tau-1$ \\
        \Xhline{1.0pt}
	\end{tabular}
	\label{tab:Resource}
\end{table*}

\subsection{Processing Schedule and Overhead Analysis}
For channel estimation under ADMA scheme, the timing of the entire design is shown in Fig. \ref{fig:timeoptimistic}, where $T_s$ is the clock cycle. we can see that each module is processed in pipeline except the UL-grouping module. The timing of the optimized architecture without rotation is the same with it is shown in Fig. \ref{fig:timeoptimistic}. The resource statistics of each module is listed in Table \ref{tab:Resource}. In addition, the latency and processing time of each module is listed in Table \ref{tab:latency}. Here, ``Latency'' is associated with one data package, and ``Processing time'' is associated with $M$ data packages. Notice that $P$ is an integer between $0$ and $M-1$ which is determined by the spatial signature of each user.

\begin{table}[htbp]
	\centering
	\caption{Latency and Processing Time}
	\begin{tabular}{ l  l  l }
        \Xhline{1.0pt}
		Modules & Latency ($T_s$) & Processing time ($T_s$) \\
		\hline
		\rowcolor{mygray}
		LS-based Estimation & $L-1$ & $L+M$ \\
		FFT & $M-1$ & $2M-1$ \\
		\rowcolor{mygray}
		Max-Selection & $M$ & $M$ \\
		Sorting & - & $\left.{\log_2 K(\log_2 K + 1)}/{2}\right.$ \\
		\rowcolor{mygray}
		Grouping & - & $K+\tau$ \\
        Extraction & $P$ & $P$ \\
		\rowcolor{mygray}
		IFFT & $\tau$ & $M+\tau$ \\
        \Xhline{1.0pt}
	\end{tabular}
	\label{tab:latency}
\end{table}

\subsection{FPGA Implementation Results}
In order to demonstrate the advantage of channel estimation under ADMA scheme, our architectures are implemented with Xilinx Virtex-7 Ultrascale vu440-flga2892-2-e FPGA. For the ease of Implementation, the parameters are set as $M=128,~K=16,~L=4,~\tau=4,~{\theta_k} \in \{-48.59^{\circ}, -14.48^{\circ}, 48.59^{\circ}, 14.48^{\circ}\}~and~\Delta \theta_k = 2^{\circ}$. The resources overhead and maximum frequency are shown in Table \ref{tab:FPGA}. We can see that the omission of rotation operations brings us $54\%$ reduction in LUTs, $57\%$ reduction in registers, $55\%$ reduction in block RAMs and $60\%$ reduction in DSPs. And for the timing constraints, since the critical path lies in the FFT module, the maximum frequency of these two architecture can both reach 217.39 megahertz.

\begin{table}[htbp]
	\centering
	\caption{FPGA Implementation Results}
	\begin{tabular}{ l  l  l }
        \Xhline{1.0pt}
		Structures & With Rotation & Without Rotation \\
		\hline
		\rowcolor{mygray}
		LUTs & $52,416$ & $24,130$ \\
		Registers & $90,191$ & $38,464$ \\
		\rowcolor{mygray}
		Block RAMs & $220$ & $100$ \\
		DSPs & $1,092$ & $432$ \\
		\rowcolor{mygray}
		Frequency (MHz) & $217.39$ & $217.39$  \\
        \Xhline{1.0pt}
	\end{tabular}
	\label{tab:FPGA}
\end{table}

\section{Conclusions}\label{sec:con}
In this paper, the hardware-efficient channel estimator based on ADMA scheme is first proposed. The corresponding optimizations on quantization and approximation are presented as well. To achieve high efficiency and low complexity, the pipelining technique and systolic structure have been employed to tailor the architecture for regularity. Finally, FPGA implementations are given. Suggestions on the choice of rotation are listed. Future work will be directed towards its application in our 5G Cloud Testbed.

\footnotesize
\bibliographystyle{IEEEtran}
\bibliography{IEEEabrv,mybib}

\begin{thebibliography}{10}
\providecommand{\url}[1]{#1}
\csname url@samestyle\endcsname
\providecommand{\newblock}{\relax}
\providecommand{\bibinfo}[2]{#2}
\providecommand{\BIBentrySTDinterwordspacing}{\spaceskip=0pt\relax}
\providecommand{\BIBentryALTinterwordstretchfactor}{4}
\providecommand{\BIBentryALTinterwordspacing}{\spaceskip=\fontdimen2\font plus
\BIBentryALTinterwordstretchfactor\fontdimen3\font minus
  \fontdimen4\font\relax}
\providecommand{\BIBforeignlanguage}[2]{{%
\expandafter\ifx\csname l@#1\endcsname\relax
\typeout{** WARNING: IEEEtran.bst: No hyphenation pattern has been}%
\typeout{** loaded for the language `#1'. Using the pattern for}%
\typeout{** the default language instead.}%
\else
\language=\csname l@#1\endcsname
\fi
#2}}
\providecommand{\BIBdecl}{\relax}
\BIBdecl

\bibitem{liu2017vlsi}
X.~Liu, H.~Xie, J.~Sha, F.~Gao, S.~Jin, X.~You, and C.~Zhang, ``{The VLSI
  architecture for channel estimation based on ADMA},'' in \emph{Proc. IEEE
  International Conference on ASIC (ASICON)}, 2017, pp. 1073--1076.

\bibitem{andrews2014will}
J.~G. Andrews, S.~Buzzi, W.~Choi, S.~V. Hanly, A.~Lozano, A.~C. Soong, and
  J.~C. Zhang, ``{What will 5G be?}'' \emph{{IEEE} J. Sel. Areas Commun.},
  vol.~32, no.~6, pp. 1065--1082, 2014.

\bibitem{shafi20175g}
M.~Shafi, A.~F. Molisch, P.~J. Smith, T.~Haustein, P.~Zhu, P.~De~Silva,
  F.~Tufvesson, A.~Benjebbour, and G.~Wunder, ``{5G: A tutorial overview of
  standards, trials, challenges, deployment, and practice},'' \emph{{IEEE} J.
  Sel. Areas Commun.}, vol.~35, no.~6, pp. 1201--1221, 2017.

\bibitem{rusek2013scaling}
F.~Rusek, D.~Persson, B.~K. Lau, E.~G. Larsson, T.~L. Marzetta, O.~Edfors, and
  F.~Tufvesson, ``{Scaling up MIMO: Opportunities and challenges with very
  large arrays},'' \emph{IEEE signal processing magazine}, vol.~30, no.~1, pp.
  40--60, 2013.

\bibitem{boccardi2014five}
F.~Boccardi, R.~W. Heath, A.~Lozano, T.~L. Marzetta, and P.~Popovski, ``Five
  disruptive technology directions for {5G},'' \emph{{IEEE} Commun. Mag.},
  vol.~52, no.~2, pp. 74--80, 2014.

\bibitem{larsson2014massive}
E.~G. Larsson, O.~Edfors, F.~Tufvesson, and T.~L. Marzetta, ``{Massive MIMO for
  next generation wireless systems},'' \emph{{IEEE} Commun. Mag.}, vol.~52,
  no.~2, pp. 186--195, 2014.

\bibitem{xie2016overview}
H.~Xie, F.~Gao, and S.~Jin, ``An overview of low-rank channel estimation for
  massive {MIMO} systems,'' \emph{IEEE Access}, vol.~4, pp. 7313--7321, 2016.

\bibitem{marzetta2010noncooperative}
T.~L. Marzetta, ``Noncooperative cellular wireless with unlimited numbers of
  base station antennas,'' \emph{{IEEE} Trans. Wireless Commun.}, vol.~9,
  no.~11, pp. 3590--3600, 2010.

\bibitem{jose2009pilot}
J.~Jose, A.~Ashikhmin, T.~L. Marzetta, and S.~Vishwanath, ``{Pilot
  contamination problem in multi-cell TDD systems},'' in \emph{Information
  Theory, 2009. ISIT 2009. IEEE International Symposium on}.\hskip 1em plus
  0.5em minus 0.4em\relax IEEE, 2009, pp. 2184--2188.

\bibitem{fernandes2013inter}
F.~Fernandes, A.~Ashikhmin, and T.~L. Marzetta, ``{Inter-cell interference in
  noncooperative TDD large scale antenna systems},'' \emph{{IEEE} J. Sel. Areas
  Commun.}, vol.~31, no.~2, pp. 192--201, 2013.

\bibitem{you2015pilot}
L.~You, X.~Gao, X.-G. Xia, N.~Ma, and Y.~Peng, ``{Pilot reuse for massive MIMO
  transmission over spatially correlated Rayleigh fading channels},''
  \emph{{IEEE} Trans. Wireless Commun.}, vol.~14, no.~6, pp. 3352--3366, 2015.

\bibitem{burr2003capacity}
A.~G. Burr, ``{Capacity bounds and estimates for the finite scatterers MIMO
  wireless channel},'' \emph{{IEEE} J. Sel. Areas Commun.}, vol.~21, no.~5, pp.
  812--818, 2003.

\bibitem{gao2016energy}
X.~Gao, L.~Dai, S.~Han, I.~Chih-Lin, and R.~W. Heath, ``{Energy-efficient
  hybrid analog and digital precoding for mmWave MIMO systems with large
  antenna arrays},'' \emph{{IEEE} J. Sel. Areas Commun.}, vol.~34, no.~4, pp.
  998--1009, 2016.

\bibitem{adhikary2013joint}
A.~Adhikary, J.~Nam, J.-Y. Ahn, and G.~Caire, ``{Joint spatial division and
  multiplexing¡ªThe large-scale array regime},'' \emph{{IEEE} Trans. Inf.
  Theory}, vol.~59, no.~10, pp. 6441--6463, 2013.

\bibitem{nguyen2013compressive}
S.~L.~H. Nguyen and A.~Ghrayeb, ``{Compressive sensing-based channel estimation
  for massive multiuser MIMO systems},'' in \emph{Proc. IEEE Wireless
  Communications and Networking Conference (WCNC)}.\hskip 1em plus 0.5em minus
  0.4em\relax IEEE, 2013, pp. 2890--2895.

\bibitem{rao2014distributed}
X.~Rao and V.~K. Lau, ``Distributed compressive {CSIT} estimation and feedback
  for {FDD} multi-user massive {MIMO} systems,'' \emph{{IEEE} Trans. Signal
  Process.}, vol.~62, no.~12, pp. 3261--3271, 2014.

\bibitem{sun2015beam}
C.~Sun, X.~Gao, S.~Jin, M.~Matthaiou, Z.~Ding, and C.~Xiao, ``Beam division
  multiple access transmission for massive {MIMO} communications,''
  \emph{{IEEE} Trans. Commun.}, vol.~63, no.~6, pp. 2170--2184, 2015.

\bibitem{fang2017low}
J.~Fang, X.~Li, H.~Li, and F.~Gao, ``{Low-rank covariance-assisted downlink
  training and channel estimation for FDD massive MIMO systems},'' \emph{{IEEE}
  Trans. Wireless Commun.}, vol.~16, no.~3, pp. 1935--1947, 2017.

\bibitem{xie2017unified}
H.~Xie, F.~Gao, S.~Zhang, and S.~Jin, ``A unified transmission strategy for
  {TDD/FDD} massive {MIMO} systems with spatial basis expansion model,''
  \emph{{IEEE} Trans. Veh. Technol.}, vol.~66, no.~4, pp. 3170--3184, 2017.

\bibitem{urquhart1984systolic}
R.~Urquhart and D.~Wood, ``Systolic matrix and vector multiplication methods
  for signal processing,'' in \emph{Proc. Insr. Elec. Eng.}, vol. 131, no.~6,
  1984, pp. 623--631.

\bibitem{ayinala2012pipelined}
M.~Ayinala, M.~Brown, and K.~K. Parhi, ``Pipelined parallel {FFT} architectures
  via folding transformation,'' \emph{{IEEE} Trans. {VLSI} Syst.}, vol.~20,
  no.~6, pp. 1068--1081, 2012.

\bibitem{cheng2007high}
C.~Cheng and K.~K. Parhi, ``{High-throughput VLSI architecture for FFT
  computation},'' \emph{{IEEE} Trans. Circuits Syst. {II}}, vol.~54, no.~10,
  pp. 863--867, 2007.

\bibitem{chang2008efficient}
Y.-N. Chang, ``{An efficient VLSI architecture for normal I/O order pipeline
  FFT design},'' \emph{{IEEE} Trans. Circuits Syst. {II}}, vol.~55, no.~12, pp.
  1234--1238, 2008.

\bibitem{batcher1968sorting}
K.~E. Batcher, ``Sorting networks and their applications,'' in \emph{Proc.
  AFIPS Spring Joint Comput. Conf}, vol.~32, 1968, pp. 307--314.

\end{thebibliography}

\end{document}